\documentclass[rmp,twocolumn,groupedaddress,showpacs,floatfix]{revtex4}

\begin{document}
\bibliographystyle{apsrev}

\title{Photons and electrons as emergent phenomena
}

\author{Michael Levin}
\author{Xiao-Gang Wen}
\homepage{http://dao.mit.edu/~wen}
\affiliation{Department of Physics, Massachusetts Institute of Technology,
Cambridge, Massachusetts 02139}
\date{May, 2005}

\begin{abstract}
Recent advances in condensed matter theory have revealed that new and exotic
phases of matter can exist in spin models (or more precisely, local bosonic
models) via a simple physical mechanism, known as "string-net condensation."
These new phases of matter have the unusual property that their collective
excitations are gauge bosons and fermions. In some cases, the collective
excitations can behave just like the photons, electrons, gluons, and quarks in
our vacuum. This suggests that photons, electrons, and other elementary
particles may have a unified origin --
string-net condensation in our vacuum. In
addition, the string-net picture indicates how to make artificial photons,
artificial electrons, and artificial quarks and gluons in condensed matter
systems.
\end{abstract}

\maketitle

\section{Introduction}

Throughout history, people have attempted to understand the universe by 
dividing matter into smaller and smaller pieces. This approach has proven 
extremely fruitful: successively smaller distance scales have revealed 
successively simpler and more fundamental structures. At the turn of the 
century, chemists discovered that all matter was formed out of a few dozen 
different kinds of particles -- atoms. Later, it was realized that atoms 
themselves were composed out of even smaller particles -- electrons, protons and
neutrons. Today, the most fundamental particles known are photons, electrons, 
quarks and a few other particles. These particles are described by a field 
theory known as the $U(1)\times SU(2)\times SU(3)$ standard model 
\cite{CheL91}. 
 
It is natural to wonder -- are photons, electrons, and quarks truly elementary? 
Or are they composed out of even smaller and more fundamental objects (perhaps 
superstrings \cite{GreSW88})? A great deal of research has been devoted to 
answering these questions.

However, the questions themselves may be fundamentally flawed. They are based
on the implicit assumption that we can understand the nature of particles by 
dividing them into smaller pieces. But does this line of thinking necessarily 
make sense? There are many examples from condensed matter physics indicating 
that sometimes, this line of thinking does not make sense. 

\begin{figure}[tb]
\centerline{
\includegraphics[width=1.0in]{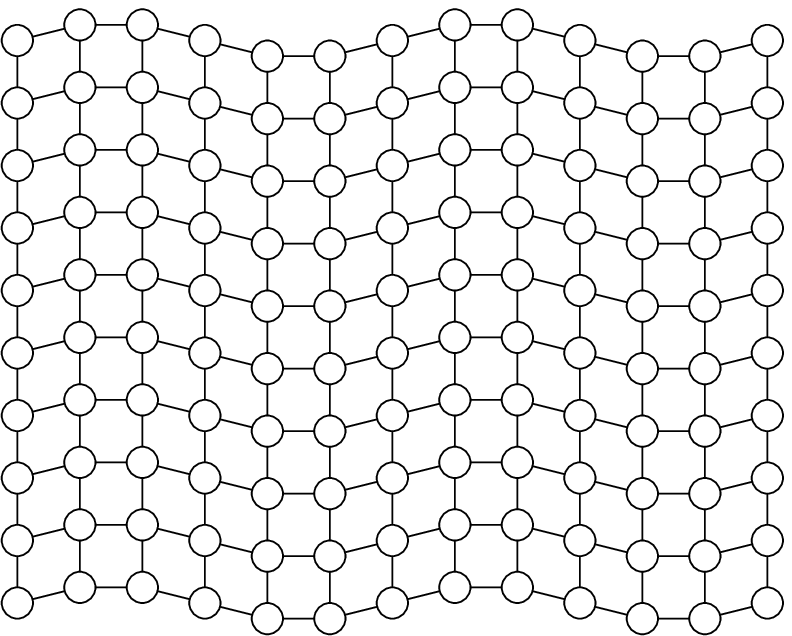}
}
\caption{
Sound waves in a crystal are particles called phonons, according 
to quantum theory.
}
\label{sound}
\end{figure}

Consider, for example, a crystal. We know that a sound wave can propagate
inside a crystal (see Fig. \ref{sound}). 
According to quantum theory, these waves behave like particles called
phonons. Phonons are no less particle-like than photons. But no one attempts
to gain a deeper understanding of phonons by dividing them into smaller
pieces. This is because phonons -- as sound waves -- are collective motions of
the atoms that form the crystal.  When we examine phonons at short distances,
we do not find small pieces that make up a phonon. We simply see the atoms in
the crystal.

This example suggests an alternate line of inquiry. Are photons, electrons, and
other elementary particles collective modes of some deeper structure?
If so, what is this ``deeper structure''? 

Ultimately, these questions will have to be answered by experiment. However, 
in this paper we would like to address the plausibility of this condensed
matter model of the universe on theoretical grounds. 

The laws of physics seem to be composed out of five fundamental ingredients:
\begin{enumerate}
\item{Identical particles}
\item{Gauge interactions}
\item{Fermi statistics}
\item{Chiral fermions}
\item{Gravity}
\end{enumerate}
The question is whether one can find a ``deeper structure'' that gives rise to 
all five of these phenomena. In addition to being consistent with our current
understanding of the universe, such a structure would be quite appealing from a
theoretical point of view: it would unify and explain the origin of these
seemingly mysterious and disconnected phenomena.

The $U(1)\times SU(2) \times SU(3)$ standard model fails to provide such a
complete story for even the first four phenomena. Although it describes
identical particles, gauge interactions, Fermi statistics and chiral fermions
in a single theory, each of these components are introduced
\emph{independently} and \emph{by hand}.  For example, field theory is
introduced to explain identical particles, vector gauge fields are
introduced to describe gauge interactions \cite{YM5491} and anticommuting
fields are introduced to explain Fermi statistics. One wonders -- where do
these mysterious gauge symmetries and anticommuting fields come from?  Why
does nature choose such peculiar things as fermions and gauge bosons to
describe itself? We hope that the ``deeper structure'' that we are looking for
can resolve these mysteries. 

So far we do not know any structure that gives rise to, and unifies all five
phenomena. In this paper, we will describe a partial solution -- a structure
that naturally gives rise to, and unifies the first three phenomena (and
possibly also the fifth \cite{Smo02}). In the language of condensed matter
physics, this structure has the unusual property that its collective modes are 
gauge bosons (such as photons) and fermions (such as electrons).

\section{Locality principle}

What kinds of ``structures'' should we look for in order to understand the
origin of gauge bosons and fermions?  In this paper, we will consider a
very general class of structures. In fact, we will impose only one
constraint on the structures we consider: locality. We will require that (a)
the total Hilbert space is a product of small Hilbert spaces 
that describe local degrees of freedom, and that (b) the Hamiltonian
only involves local interactions. For the sake of concreteness, we will also
restrict our attention to lattice models.

These two requirements naturally lead us to large class of structures that
we call ``local bosonic models'' or (generalized) ``spin models''. These are lattice 
models where each lattice site can be in a few states $|a\>$ labeled by 
$a=0,1,2,...,N$. 

The question is -- can we find a local bosonic model whose 
collective modes are fermions and gauge bosons? 

\section{From new phases of matter to a unification of gauge interactions and 
Fermi statistics}

\begin{figure}[tb]
\centerline{
\includegraphics[width=2.0in]{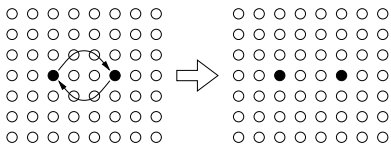}
}
\caption{
The empty circles correspond to sites in the $|0\>$ state.
The solid circles correspond to sites in the $|1\>$ state.
The states before and after the exchange are
naturally the same, and hence the $|1\>$ states represent
bosonic particles.
}
\label{latt}
\end{figure}

At first, it appears that the local bosonic models do not work. Consider, for
example, a local bosonic model whose ground state has $a = 0$ for every lattice
site. We think of this ground state as the vacuum. A particle in the vacuum
corresponds to a state with $a \neq 0$ for one site, and $a = 0$ for all other
sites (see Fig. \ref{latt}). One can easily check that these particles are
identical bosons. They are a particular kind of boson -- a scalar boson. They
are very different from gauge bosons and they are definitely not fermions.
Thus, local bosonic models with this particularly simple ground state do not
have the appropriate collective modes.

But we should not give up just yet. We know that the properties of
excitations depend on the properties of the ground state. If we change the
ground state qualitatively, we may obtain a new phase of matter with new
excitations. These new excitations may be gauge bosons or fermions.

For many years, this was thought to be impossible. This conviction was
largely based on Landau's symmetry breaking theory -- a general framework for 
describing phases of matter \cite{L3726}. According to Landau theory,
phases of matter are characterized by the symmetries of their 
ground states. The ground state symmetry directly determines the properties of 
the collective excitations \cite{LanL58}. Using Landau theory, one can show 
that the collective modes can be very different for different ground states,
but that they are always scalar bosons. There is no sign of gauge bosons or 
fermions.

After the discovery of the fractional quantum Hall effect 
(\cite{TSG8259}; \cite{L8395}), it became clear that Landau theory could not describe all
possible phases of matter. Fractional quantum Hall states contain a new kind of
order -- topological order \cite{Wtoprev,Wen04} -- that is beyond Landau theory. 
The collective excitations of the FQH states are not scalar bosons. Instead, 
they have fractional statistics \cite{ASW8422}, statistics somewhere in between
Bose and Fermi statistics \cite{LM7701,W8257}.

\begin{figure}[tb]
\centerline{
\includegraphics[width=2.0in]{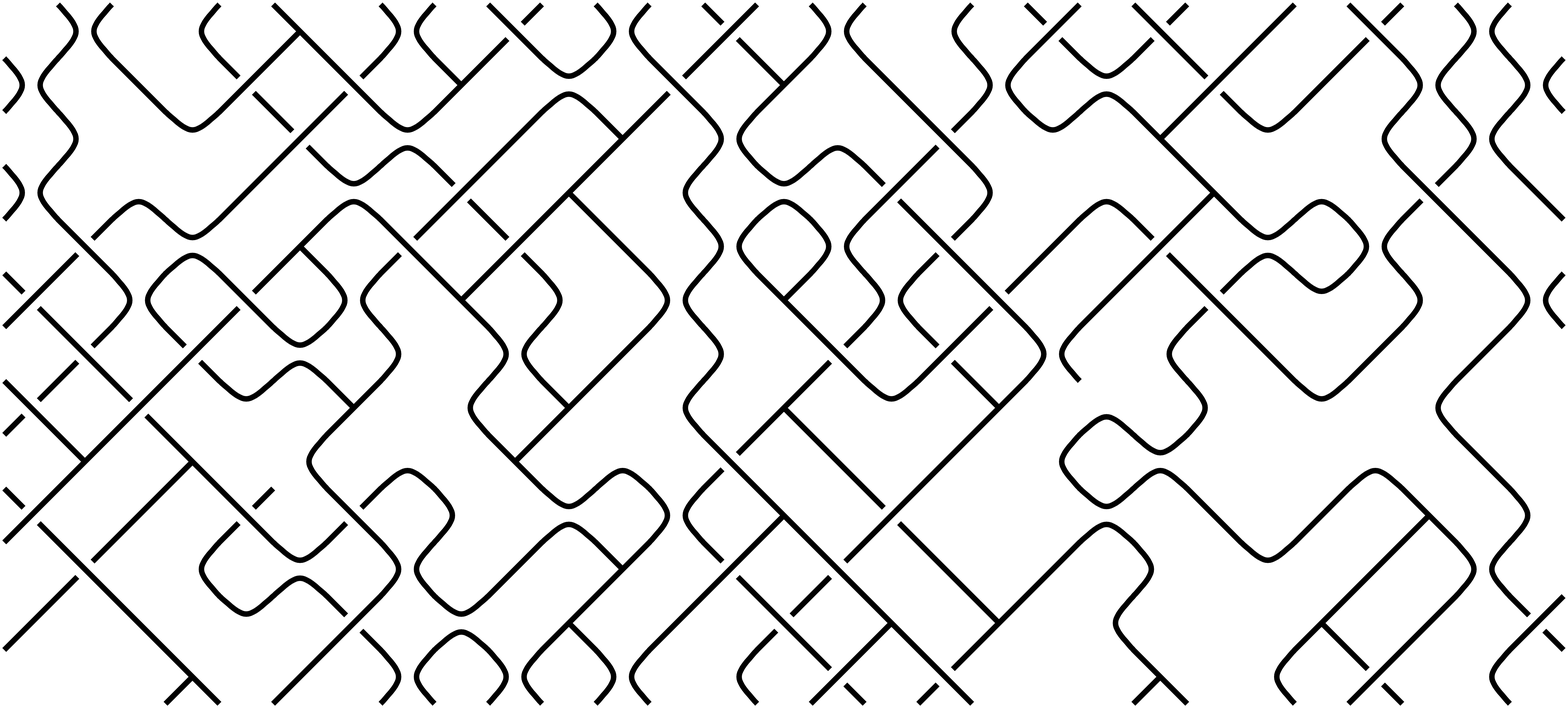}
}
\caption{
A typical string-net configuration in a string-net condensed state.
The fluctuations of the strings correspond to gauge bosons (such as photons)
and the ends of strings correspond to fermions (such as electrons).
}
\label{stringnetS}
\end{figure}

So there is still hope. Perhaps gauge bosons and fermions can emerge from new
phases of matter -- phases of matter that are beyond Landau theory.  This is
indeed the case. Recently, it was realized that a new class of phases of matter
-- 3D string-net condensed phases \cite{LWsta,Wqoem,LWstrnet} -- have the
desired property. String-net condensed states are 
liquids of fluctuating
networks of strings (see Fig. \ref{stringnetS}). In some sense, they are
analogous to Bose condensed states, except that the condensate is formed from
extended objects rather then particles. \footnote{While the terminology is 
similar, the reader should not confuse the theory of string-net condensation
with string theory. The two theories are quite different. One important 
distinction is that the strings in string theory are microscopic -
with a typical length on the order of the Planck length - while the extended 
objects in string-net condensates are macroscopic - with a typical length 
on the order of the system size.} However, the collective excitations
above string-net condensed states are not scalar bosons, but rather gauge
bosons and fermions! Roughly speaking, the vibrations of the strings give rise
to gauge bosons, while the ends of the strings correspond to fermions (see
section \ref{prop}).

This result may change our conception of gauge bosons and fermions. If we
believe that the vacuum is some kind of string-net condensed state, then gauge
bosons and fermions are just different sides of the same coin \cite{LWsta}.
In other words, string-net condensation provides a way to unify gauge bosons
(such as photons) and fermions (such as electrons). It explains what gauge
bosons and fermions are, where they come from, and why they exist. One 
application of this deeper understanding is the construction of 3D spin systems
that contain both artificial photons and artificial electrons as low energy 
collective excitations (see section \ref{3Dmodel}) \cite{Wlight,Wqoem}.


\section{String-net condensation}

What is string-net condensation? Let us first describe string-nets and
string-net models. A string-net is a network of strings. The strings, which
form the edges or links of the network, can come in different "types" and can
carry a sense of orientation. Thus, string-nets can be thought of as networks
or graphs with oriented, labeled edges. 

\begin{figure}[tb]
\centerline{
\includegraphics[scale=0.35]{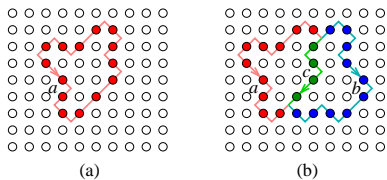}
}
\caption{
The empty circles are sites in the $|0\>$ state. The filled circles are sites
in some $|a\neq 0\>$ state. (a) A string formed by a loop of sites with state
$a$. (b) A string-net formed by a type-$a$ string, a type-$b$ string
and a type-$c$ string. The string-net has two branching points
where the three types of strings join.
}
\label{strlatt}
\end{figure}

String-net models are a special class of local bosonic models whose low 
energy physics is described by fluctuating string-nets. To
understand how this works, consider a general local bosonic model with the 
states on site $\v i$ labeled by $a_{\v i}=0,1,...,N$. The states of this model
can be thought of as configurations of string-nets in space. We regard the 
state with all $a_{\v i} = 0$ as the no-string state. We think of the state 
with a loop of sites in the $|a\neq 0\>$ state as containing a closed 
``type-$a$'' string (see Fig. \ref{strlatt}a). More complicated states will 
correspond to networks of strings as in Fig. \ref{strlatt}b. The 
orientations of the corresponding strings are determined by some specified
orientation convention, where one assigns some (arbitrary) orientation 
to each site $i$.

For most local bosonic models, this string-net picture is misleading.
Each local bosonic degree of freedom fluctuates independently and the physics
is better described by individual spins than by extended objects. However, for 
one class of local bosonic models, the string-net picture \emph{is} 
appropriate. These are local bosonic models with the property that when strings
end or change string type in empty space, the system incurs a finite energetic 
penalty. In these models, energetic constraints force the local bosonic degrees
of freedom on the lattice sites to organize into effective extended objects. 
The low energy physics is then described by the fluctuations of these effective
string-nets. String-net models are local bosonic models with this additional 
property.

\begin{figure}[tb]
\centerline{
\includegraphics[width=1.1in]{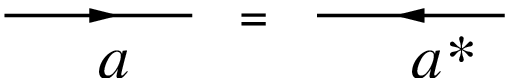}
}
\caption{
$a$ and $a^*$ label strings with opposite orientations.
}
\label{strOrt}
\end{figure}

To specify a particular string-net model, one needs to provide several pieces 
of information that characterize the structure of the effective string-nets.
First, one needs to give the number of string types $N$. Second, one needs to
specify the branching rules -- that is, what triplets of string types $(abc)$ 
can meet at a point. (Here for simplicity, we only consider the simplest type 
of branching -- where three strings join at a point). The branching rules are
specified by listing the ``legal'' branching triplets $\{(abc),(def),...\}$. 
For example, if $(abc)$ is legal then the string-net configuration shown in 
Fig. \ref{strlatt}b is allowed. Finally, one needs to describe the string
orientations: for every string type $a$, one needs to specify another
string type $a^*$ that corresponds to a string with the opposite orientation 
(see Fig. \ref{strOrt}). A string loses its sense of orientation if its string 
type satisfies $a=a^*$. 

\begin{figure}[tb]
\centerline{
\includegraphics[scale=0.35]{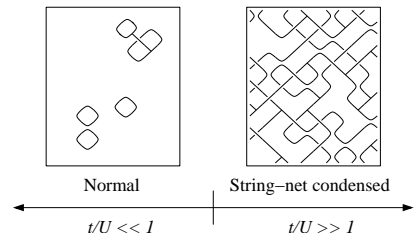}
}
\caption{
A schematic phase diagram for the generic string-net Hamiltonian (\ref{genh}).
When $t/U$ is small the system is in the normal phase. The ground state 
consists of a few small string-nets. When $t/U$ is large the
string-nets condense and large fluctuating string-nets fill all of space. We
expect a phase transition between the two states at some $t/U$ of order unity.
We have omitted string labels and orientations for the sake of clarity.
}
\label{stringnetSUt}
\end{figure}

Given a string-net model with some string types, branching rules, and string 
orientations, we can imagine writing down a Hamiltonian to describe the 
dynamics of the string-nets. A typical string-net Hamiltonian $H$ is a sum of 
potential and kinetic energy pieces and a constraint term:
\begin{equation}
H = U H_{U} + t H_{t} + V H_c
\label{genh}
\end{equation}
The constraint term $H_c$ enforces the branching roles by making ``illegal''
branching points cost a huge energy $V$. Because of this term, the low energy 
states contain only ``legal'' branchings. The kinetic energy $H_{t}$ gives 
dynamics to these low energy string-net states while the potential
energy $H_{U}$ is typically some kind of string tension. When $U \gg t$, the
string tension dominates and we expect the ground state to be the no-string
state with a few small string-nets. On the other hand, when $t \gg U$, the
kinetic energy dominates, and we expect the ground state to consist of many
fluctuating string-nets (see Fig. \ref{stringnetSUt}). Large string-nets with a
typical length on the order of the system size fill all of space. We expect
that there is a quantum phase transition between the two states at some $t/U$
on the order of unity. Because of the analogy with particle condensation, we
say that the large $t$, highly fluctuating string-net phase is ``string-net
condensed.''

\section{Wave functions for string-net condensates}

String-net condensed phases are new phases of matter with many
interesting properties (\cite{Walight}; \cite{FNS0311}; \cite{LWstrnet}). But how can we describe them quantitatively?
One approach is to write down a ground state string-net wave function 
$\Phi(\text{string-nets})$. However, string-net condensed wave functions
are usually too complicated to write down explicitly. Therefore, we will use a
more indirect approach: we will describe a series of local constraint equations
on string-net wave functions which have a unique solution $\Phi$. In this way, 
we can construct potentially complicated string-net wave functions without 
writing them down explicitly.

Before we state the constraint equations, we note that we can project a
three-dimensional string-net configuration onto a two dimensional plane, 
resulting in a two-dimensional graph with branching and crossings (see Fig. 
\ref{stringnetS}). Thus, a wave function of three-dimensional string-nets can 
also be viewed as a wave function of the projected two-dimensional graphs.

\begin{figure}[tb]
\centerline{
\includegraphics[width=1.7in]{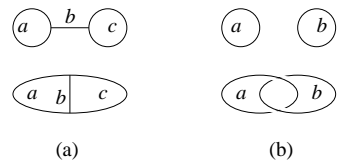}
}
\caption{
(a) Three strings with different connections. (b) Two strings with different
crossings. The numbers of the crossings are 0 and 2 respectively.
}
\label{cnnlnk}
\end{figure}

The local constraints relate the amplitudes of string-net configurations that 
only differ by small local transformations. To write down a set of these local
constraint equations or local rules, one first chooses a real tensor $d_i$ and 
two complex tensors $F^{ijm}_{kln}$, $\om^k_{ij}$ where the indices 
$i,j,k,l,m,n$ run over the different string types  $0,1,...N$. The local rules 
are then given by:
\begin{align}
\label{lclrl3}
 \Phi
\bpm \includegraphics[height=0.3in]{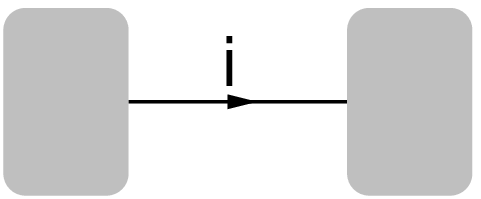} \epm  &=
\Phi 
\bpm \includegraphics[height=0.3in]{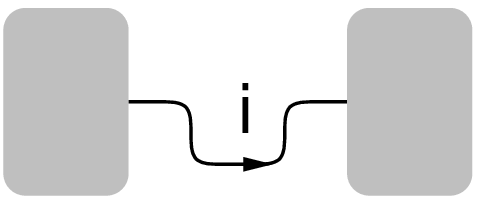} \epm
\nonumber\\
 \Phi
\bpm \includegraphics[scale=0.35]{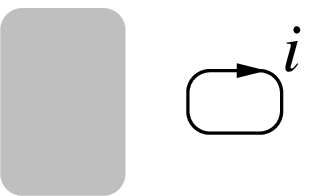} \epm  &=
d_i\Phi 
\bpm \includegraphics[scale=0.35]{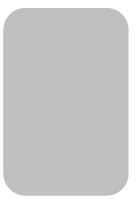} \epm
\nonumber\\
\Phi
\bpm \includegraphics[scale=0.35]{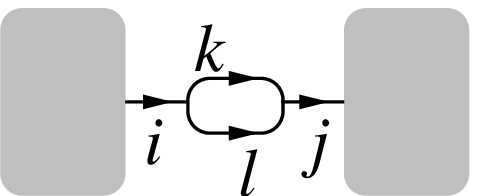} \epm  &=
\delta_{ij}
\Phi 
\bpm \includegraphics[scale=0.35]{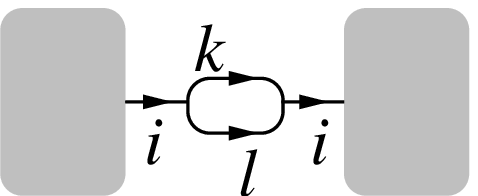} \epm
\nonumber\\
\Phi
\bpm \includegraphics[scale=0.35]{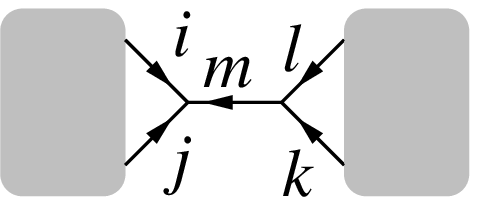} \epm  &=
\sum_{n=0}^N
F^{ijm}_{kln}
\Phi 
\bpm \includegraphics[scale=0.35]{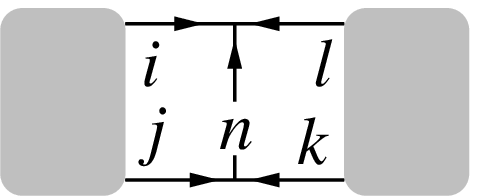} \epm
\nonumber\\
\Phi
\bpm \includegraphics[scale=0.35]{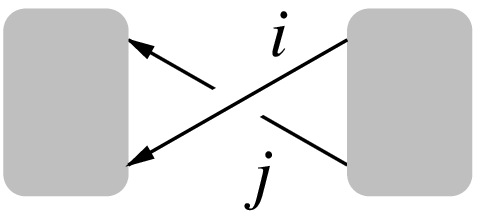}\epm &= 
\sum_{k=0}^N
\om^{k}_{ij}
\Phi
\bpm \includegraphics[scale=0.35]{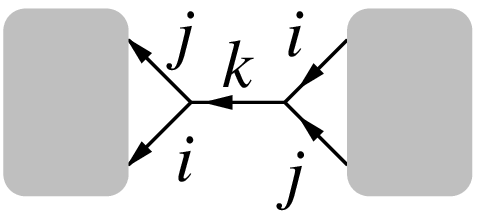}\epm
\nonumber\\
\Phi
\bpm \includegraphics[scale=0.35]{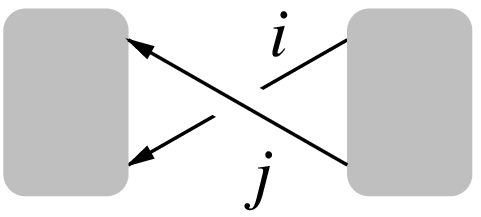}\epm &= 
\sum_{k=0}^N
\om^{k}_{ij}
\Phi\bpm\includegraphics[scale=0.35]{Brd2O.eps}\epm
\end{align}
where the shaded gray areas represent other parts of string-nets that are not
changed. Here, the type-$0$ string is interpreted as the no-string state. 
We would like to mention that we have drawn the first local rule
somewhat schematically. The more precise statement of this rule is that any two
string-net configurations that can be continuously deformed into each other 
have the same amplitude. In other words, the string-net wave function $\Phi$ 
only depends on the topologies of the projected graphs; it only depends on how 
the strings are connected and crossed (see Fig. \ref{cnnlnk}).

By applying the local rules in \Eq{lclrl3} multiple times, one can compute
the amplitude of any string-net configuration in terms of the amplitude of
the no-string configuration. Thus \Eq{lclrl3} determines the 
string-net wave function $\Phi$. \footnote{For example, we can compute the 
amplitude
\begin{align}
\Phi
\bpm \includegraphics[scale=0.35]{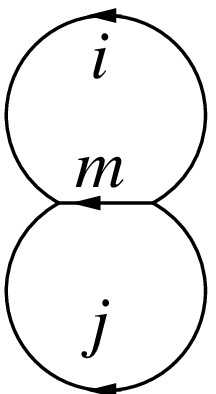} \epm  &=
\sum_{n=0}^N
F^{ijm}_{j^*i^*n}
\Phi 
\bpm \includegraphics[scale=0.35]{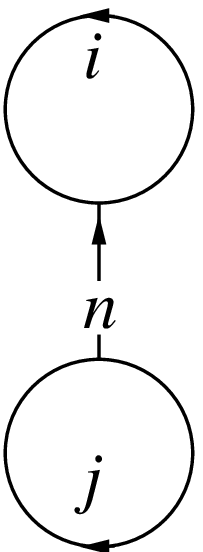} \epm =
F^{ijm}_{j^*i^*0}
\Phi 
\bpm \includegraphics[scale=0.35]{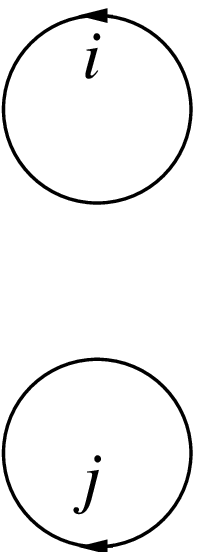} \epm \nonumber\\ &=
F^{ijm}_{j^*i^*0} d_i d_j \nonumber
\end{align}
by appying the fourth rule, the third rule, and the second rule in sequence.}

However, an arbitrary choice of $(d_i, F^{ijk}_{lmn}, \om^{k}_{ij})$ does not 
lead to a well defined $\Phi$. This is because two string-net configurations 
may be related by more then one sequence of local rules. We need to choose the 
$(d_i, F^{ijk}_{lmn}, \om^{k}_{ij})$ carefully so that different sequences of 
local rules produce the same results. That is, we need to choose 
$(d_i, F^{ijk}_{lmn}, \om^{k}_{ij})$ so that the rules are self-consistent. 
Finding these special tensors is the subject of tensor category theory 
\cite{Tur94}. It has been shown that only those that satisfy
\cite{LWstrnet}
\begin{align}
\label{pentbrdid}
F^{ijk}_{j^*i^*0} &= \frac{v_{k}}{v_{i}v_{j}} \del_{ijk} \nonumber \\ 
F^{ijm}_{kln} = F^{lkm^*}_{jin} &= F^{jim}_{lkn^*} = F^{imj}_{k^*nl}
\frac{v_{m}v_{n}}{v_{j}v_{l}} \nonumber \\
\sum_{n=0}^N
F^{mlq}_{kp^*n} F^{jip}_{mns^*} F^{js^*n}_{lkr^*}
&= F^{jip}_{q^*kr^*} F^{riq^*}_{mls^*}
\nonumber\\
\omega^{m}_{js}F^{sl^*i}_{kjm^*}\omega^{l}_{si}
\frac{v_j v_s}{v_m} &=
\sum_{n=0}^{N}F^{ji^*k}_{s^*nl^*}\omega^{n}_{sk}F^{jl^*n}_{ksm^*} 
\nonumber \\
\omega^{j}_{is} &= \sum_{k=0}^N \omega^{k}_{si^*}F^{i^*s^*k}_{isj*}
\end{align}
will result in self-consistent rules and a well defined string-net wave 
function $\Phi$. Such a wave function describes a string-net condensed state. 
Here, we have introduced some new notation: $v_i$ is defined by 
$v_i = v_{i^*} = \sqrt{d_i}$ while $\del_{ijk}$ is given by 
\begin{equation*}
\del_{ijk}=
\begin{cases}
1, & \text{if $(ijk)$ is legal,}\\
0, & \text{otherwise}
\end{cases}
\end{equation*}

There is a one-to-one correspondence between 3D string-net condensed phases and
solutions of \Eq{pentbrdid}. It is interesting to compare this with a more
familiar classification scheme: the classification of crystals. In a crystal, 
atoms organize themselves into a very regular pattern -- a lattice. Since 
different lattice structures are distinguished by their symmetries, we can use 
group theory to classify all the 230 crystals in three dimensions. In much the 
same way, string-net condensed states are highly structured. The different 
possible structures are described by solutions to \Eq{pentbrdid}. Tensor 
category theory provides a classification of the solutions of \Eq{pentbrdid}, 
which leads to a classification of string-net condensates. Thus tensor category
theory is the underlying mathematical framework for understanding string-net 
condensed phases, just as group theory is for symmetry breaking phases.

\section{Properties of collective excitations above string-net condensed
states}
\label{prop}

Both crystals and string-net condensed states contain highly organized
patterns. Fluctuations of these patterns lead to collective excitations.  We
know that the fluctuations of the lattice pattern are phonons. But what are the
fluctuations of the pattern of string-net condensation? It turns out that the
collective excitations above string-net condensed states are gauge bosons
(\cite{KS7595}; \cite{BMK7793}; \cite{F7987}; \cite{FNN8035}; \cite{S8067}) and
fermions \cite{LWsta}.  The gauge bosons correspond to vibrations of the
string-nets while the fermions correspond to the ends of strings. 

\begin{figure}[tb]
\centerline{
\includegraphics[scale=0.35]{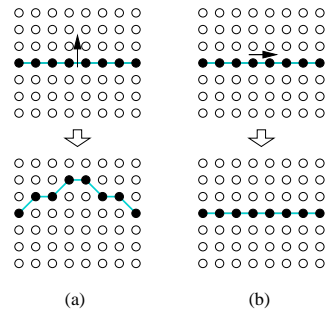}
}
\caption{
(a) A transverse motion of a string results in a new state and leads to
a collective excitation.
(b) 
A motion along the string does not result in any new states. Such a motion
does not lead to any collective excitations.
}
\label{strvib}
\end{figure}

Physically, the vibrating-string picture of gauge boson makes a lot of sense.
We know that atoms in a crystal can vibrate in three directions and that this
leads to three phonon modes. In contrast, strings can only vibrate in two
transverse directions (see Fig.  \ref{strvib}). So string vibrations can only
produce excitations with two modes. This explains why gauge bosons (such as
photons) have only two transverse polarizations.

There are many different gauge theories, each associated with a different
gauge group and a different kind of gauge boson. (For example, the gauge
group for electromagnetism is $U(1)$). Hence, it is natural to 
wonder -- what is the gauge group associated with each string-net condensate? 
It turns out that the gauge group is determined by the same data
$(d_i, F^{ijk}_{lmn}, \om^{k}_{ij})$ that characterizes the condensate.

Given a gauge group $G$, the corresponding string types, branching rules and
 $(d_i,F^{ijk}_{lmn}, \om^{k}_{ij})$ are determined as follows. The number 
of string types $N+1$ is given by the number of irreducible representations of 
$G$; each string type $i$ corresponds to a representation. The branching rules
are the Clebsch-Gordan rules for $G$; that is, $(abc)$ is a ``legal'' branching
if and only if the tensor product of the corresponding representations $a$, 
$b$, $c$ contains the trivial representation. The $d_i$ are the dimensions
of the irreducible representations $i$ and the tensor $F^{ijk}_{lmn}$ is the
$6j$ symbol of the group $G$. Finally, the tensor $\om^k_{ij}$ is given by
$\om^k_{ij} = -\frac{v_k}{v_iv_j}$ if $i = j$ and the invariant tensor
in the tensor product $i \otimes i \otimes k^*$ is antisymmetric, and
$\om^k_{ij} = \frac{v_k}{v_iv_j}$ otherwise. For any group $G$, this 
construction provides a solution  $(d_i,F^{ijk}_{lmn}, \om^{k}_{ij})$ to 
\Eq{pentbrdid}. Therefore, string-net condensed states can generate gauge 
bosons with any gauge group.

The second type of excitation of string-net condensed states are point
defects in the condensate. These can be created by adding an open string to the
condensate: new defects are formed at the ends of the string. 

\begin{figure}[tb]
\centerline{
\includegraphics[scale=0.35]{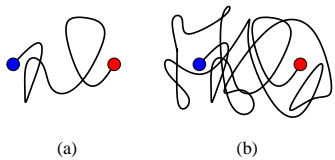}
}
\caption{
(a) An open string with two ends. 
(b) The same open string is unobservable when
placed in a
background of string-net condensed state. 
Thus the ends of open strings behave
like independent particles.
}
\label{strunob}
\end{figure}

These defects behave like independent particles even though they are the
endpoints of open strings. This is because the string connecting the two ends
blends in with the condensed string already in the ground state and hence is
unobservable (see Fig. \ref{strunob}). 
Only the endpoints of the string stand out and are observable.
Thus the ends of strings behave like point-like objects and can be treated as 
particles.

It turns out that the string endpoints interact with the string vibrations
just like charges interact with gauge bosons. Thus the endpoints of open
strings are the charges of the gauge theory. For example, if the vibrations of
the strings behave like photons, then the endpoints of the strings behave like
electric charges.

For some string-net condensates the ends are bosons while for others the ends
are fermions. What determines the statistics of the charges? It turns out that
the statistics are also determined by the $(d_i, F^{ijk}_{lmn}, \om^{k}_{ij})$
associated with the condensate.  To see this, we note that
$\Phi \bpm \includegraphics[scale=0.3]{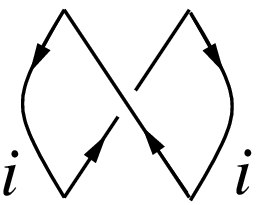} \epm$ is the amplitude to
create two pairs of particle-hole, then exchange the particles, and then
annihilate the particle-hole pairs.  So the phase of the exchange is the phase
of $ \Phi \bpm \includegraphics[scale=0.3]{2phex.eps} \epm $ which turns out to
be $e^{i\th}=\om^0_{i^*i} d_i$. Thus the end of a type-$i$ string is a fermion
if $\om^{0}_{i^*i}d_i =-1$ and a boson if $\om^{0}_{i^*i}d_i =1$.  We note that
there is no way to create a single end of a string all by itself. Thus the
string-net picture of fermions explains why we cannot create a single fermion.

\section{Simple examples of string-net condensed states}

\subsection{$Z_2$ gauge theory}

The simplest string-net model contains only one type of string ($N = 1$), and 
has no branching. In this case, one finds that \Eq{pentbrdid} has two 
solutions. Each solution corresponds to a set of self-consistent local rules. 
The two sets of local rules, labeled by $\eta=\pm 1$, are given by
\begin{align}
\Phi
\bpm \includegraphics[height=0.17in]{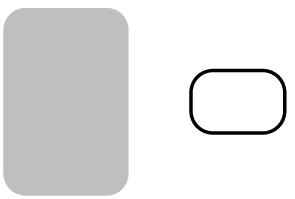} \epm  &=
\Phi 
\bpm \includegraphics[height=0.17in]{X0.eps} \epm , \ \ \ \ \ 
 \Phi
\bpm \includegraphics[height=0.17in]{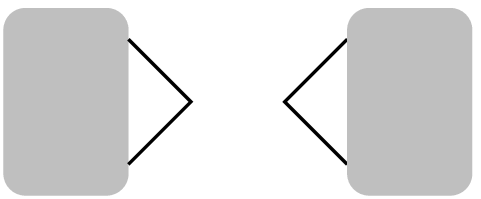} \epm  =
\Phi 
\bpm \includegraphics[height=0.17in]{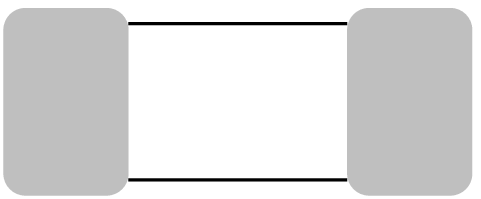} \epm,
\nonumber\\
\Phi
\bpm\includegraphics[height=0.17in]{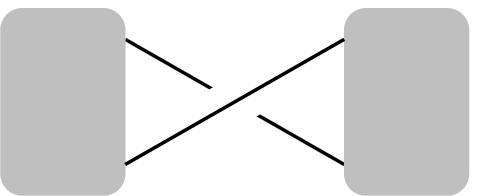}\epm
&= \Phi
\bpm\includegraphics[height=0.17in]{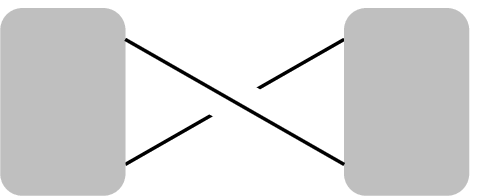}\epm
= 
\eta \Phi
\bpm\includegraphics[height=0.17in]{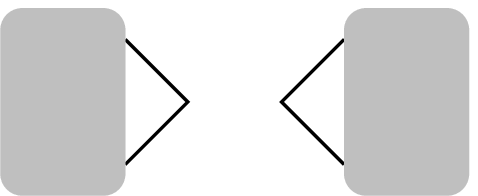}\epm .
\end{align}
The local rules are so simple that we can calculate
the corresponding string-net wave function explicitly. We find
$ \Phi(X) = \eta^{X_c}$,
where $X_c$ is the number of the crossings in the string-net configuration
$X$ (see Fig. \ref{cnnlnk}b). The two string-net wave functions correspond to 
two different string-net condensed phases. In the $\eta = +1$ phase,
the string fluctuations above the condensate are described by a $Z_2$ gauge 
theory. The ends of the strings are bosonic $Z_2$ gauge charges. In the
$\eta = -1$ phase, the string fluctuations are still described by a $Z_2$ gauge
theory, but the ends of the strings are fermions.

\subsection{$U(1)$ gauge theory with fermions}
\label{3Dmodel}

\begin{figure}[tb]
\centerline{
\includegraphics[width=0.7in]{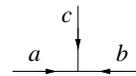}
}
\caption{
The branching rule $a+b+c=0$.
}
\label{brnU1}
\end{figure}

To construct a string-net condensate with photon-like and electron-like 
excitations, we need a string-net model with oriented stings labeled by 
integers $a=0,\pm1,\pm2,...$. We need the following branching rules: $(abc)$ is
legal if $a+b+c=0$ (see Fig. \ref{brnU1}). These branching rules have a simple
physical interpretation if we view the strings as electric flux lines and 
the labels $a$ as measuring the amount of electric flux flowing through the 
string. The branching rule $a+b+c=0$ is then simply a statement of flux 
conservation (e.g. Gauss' law).

One finds that \Eq{pentbrdid} has two solutions. One of these solutions
can be represented by the following local rules:
\begin{align*}
\Phi
\bpm \includegraphics[scale=0.35]{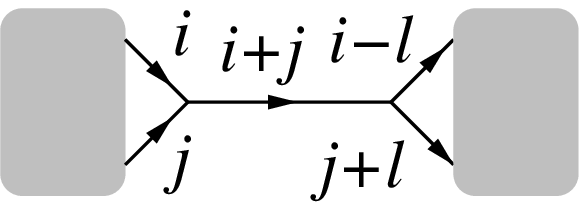} \epm  &=
\Phi 
\bpm \includegraphics[scale=0.35]{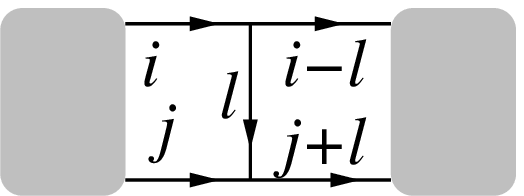} \epm  ,
\nonumber\\
\Phi
\bpm\includegraphics[scale=0.35]{Brd1O.eps}\epm
&= 
\Phi
\bpm\includegraphics[scale=0.35]{Brd1pO.eps}\epm
\nonumber\\
&=
(-1)^{i\times j} \Phi
\bpm\includegraphics[scale=0.35]{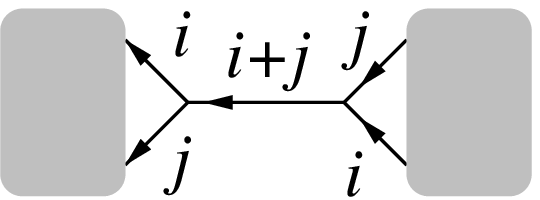}\epm .
\end{align*}
The local rules lead to the string-net wave function $\Phi(X) = (-1)^{X_{co}}$,
where $X_{co}$ is the number of the crossings between strings labeled by
\emph{odd} integers, in the string-net configuration $X$.

The collective excitations in the above string-net condensed phase are $U(1)$
gauge bosons which behave in every way like the photons in our vacuum. We call
these excitations ``artificial photons.'' The ends of type-$1$ strings behave 
like fermions with unit charge. They interact with artificial photons in the 
same way that electrons interact with photons. Therefore, we call the ends of 
type-$1$ strings ``artificial electrons.'' More generally, the ends of type-$i$
strings behave like bound states of $i$ artificial electrons.

\section{Artificial photons and Artificial electrons}

We have seen that, for any solution $(d_i, F^{ijk}_{lmn}, \om^{ik}_{j})$ of
\Eq{pentbrdid}, we can construct a corresponding string-net condensed state. 
The properties of collective excitations of this state are determined by the 
data $(d_i,F^{ijk}_{lmn}, \om^{ik}_{j})$. Now the question is, can we realize 
such a string-net condensed state in a condensed matter system? The answer is 
yes, at least theoretically. It was shown recently that for every solution 
$(d_i, F^{ijk}_{lmn}, \om^{ik}_{j})$ of \Eq{pentbrdid}, we can construct an 
exactly soluble local bosonic model such that the ground state of the model is 
the corresponding string-net condensed state \cite{LWstrnet}. 
The collective excitations in such a model are the gauge bosons and fermions 
discussed above. So in principle, we can construct condensed matter systems 
which can generate gauge bosons with arbitrary gauge groups and fermions with
arbitrary gauge charges.

However, these exactly soluble bosonic models are usually complicated
and hard to realize in real materials. On the other hand, if we only want to
make artificial photons, then there is a simple spin-$S$ model on the 
(three-dimensional) pyrochlore lattice.\footnote{The pyrochlore lattice is 
a three-dimensional network of corner-sharing tetrahedra. One way to
obtain the pyrochlore lattice is to place lattice sites at the midpoints of the
bonds of the diamond lattice.} The Hamiltonian is given by
\begin{equation}
H=
J_1\sum_{\v i} (S^z_{\v i})^2
+J_2\sum_{\<\v i\v j\>} S^z_{\v i} S^z_{\v j}
+J_\perp \sum_{\<\v i\v j\>,a=x,y}
S^a_{\v i} S^a_{\v j},
\label{artph}
\end{equation}
where $\<\v i\v j\>$ are nearest neighbors. It turns out that the above model 
exhibits string-net condensation for integer $S$ and 
$J_1 \gg |J_{\perp}^3/J_1^2| \gg |J_1 - J_2|$ \cite{Walight}. In this limit,
the model contains gapless artificial photons as its low energy excitations. 
The ground state -- a string-net condensed state -- represents a new state of 
matter that cannot be described by Landau's symmetry breaking theory. A similar
model with spin $S = 1/2$ may also contain artificial photons \cite{HFB0404}.

To understand this result and its generalizations to other lattices, it is 
useful to consider the low energy behavior of (\ref{artph}). In the limit 
of large $J_1 \approx J_2$, the above model has a low energy sector consisting 
of states satisfying 
$S_{t1}^z + S_{t2}^z + S_{t3}^z + S_{t4}^z = 0$ for all tetrahedra $t$ in the 
pyrochlore lattice. Restricting to this subspace -- which can be thought
of as the string-net sector -- we find that the low energy effective 
Hamiltonian is given by 
\begin{align}
\label{eff}
H_{eff} &= 
g \sum_{\v p} ( B_{\v p}+ 
h.c.) + \del J\sum_{\v i} (S^z_{\v i})^2 , \\
B_{\v p}&\equiv S_{1}^{+}S_{2}^{-}S_{3}^{+}S_{4}^{-}S_{5}^{+}S_{6}^{-} .
\nonumber
\end{align}
Here the sum runs over the hexagonal plaquettes $\v p$ of the pyrochlore
lattice, and $1,\cdots,6$ label the sites of the hexagon $\v p$. The
two coupling constants are given by $\del J =  J_1-J_2$ and 
$g = \frac{3J_{\perp}^3}{2J_2^2}$. 

According to the string picture, the $\del J$ term corresponds to a string 
tension term, while the $g$-term corresponds to a string kinetic energy term. 
When $|g| \gg |\del J|$, the string fluctuations overwhelm the string tension,
and the string-nets condense. The result is a new state of matter with
gapless artificial photons as its low energy excitations.
  
\begin{figure}[tb]
\centerline{
\includegraphics[scale=0.8]{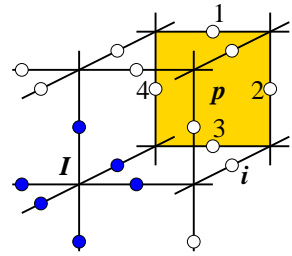}
}
\caption{
A picture of the cubic lattice model (\ref{strnetH}) with artificial photons. 
The term $Q_{\v I} = (-1)^{\v I}\sum_{\text{legs of }\v I} S_{\v i}^z$ acts on 
the six ``legs'' of $\v I$ -- that is, the six spins adjacent to $\v I$, drawn 
above as filled dots. The term $B_{\v p}= S^+_{1} S^-_{2} S^+_{3} S^-_{4}$ acts
on the four spins, labeled by 1,2,3,4, along the boundary of the plaquette 
$\v p$.
}
\label{cubbos}
\end{figure}

With this understanding, one can easily generalize this result to other
lattices. A particularly simple example is a cubic lattice model with
spins on the links \cite{LWtoap}. The Hamiltonian is given by  
\begin{align}
\label{strnetH}
H &= V\sum_{\v I}  Q^2_{\v I}
+ g \sum_{\v p} (B_{\v p}+h.c.)
+ \del J \sum_{\v i} (S^z_{\v i})^2 ,
\\
B_{\v p} &= S^+_{1} S^-_{2} S^+_{3} S^-_{4},\ \ \ \ \ \ \ \ \ 
Q_{\v I} = (-1)^{\v I}\sum_{\text{legs of }\v I} S_{\v i}^z ,
\nonumber 
\end{align}
where $\v I=(I_x, I_y, I_z)$ labels the vertices, $\v i$ labels the links and 
$\v p$ labels the plaquettes of the cubic lattice (see Fig. \ref{cubbos}). 
Also, $(-1)^{\v I} \equiv (-1)^{I_x+I_y+I_z}$.

As in the previous case, the model (\ref{strnetH}) has a low energy sector 
consisting of string-net states, in the limit of large $V$. When 
$|g| \gg |\del J|$, the string-nets condense giving rise to a phase 
with gapless artificial photons as its low energy excitations.

\begin{figure}[tb]
\centerline{
\includegraphics[scale=0.8]{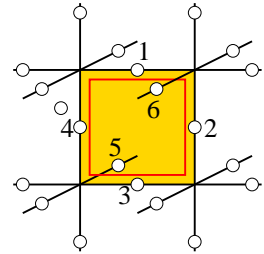}
}
\caption{
A picture of the modified cubic lattice model (\ref{twistedH}) with artificial
photons and artificial electrons. The term
$\tilde{B}_{\v p} = S^+_{1} S^-_{2} S^+_{3} S^-_{4}(-1)^{S^z_5+S^z_6}$
acts on the four spins, labeled by 1,2,3,4, along the boundary of the 
plaquette $\v p$, and the two spins, labeled by 5,6 on the ``crossed'' links 
adjacent to $\v p$. The ``crossed'' links are defined by projecting the cubic 
lattice onto the plane and drawing a small closed curve just inside but along 
the boundary of each plaquette $\v p$. The edges adjacent to $\v p$ that cross
this closed curve, are known as ``crossed'' links.
}
\label{cubferm}
\end{figure}

In the above two models, the electric charges are bosonic. However, one can
obtain models with fermionic electric charges (e.g. artificial electrons) by
modifying these Hamiltonians in a simple way: one simply multiplies the ring 
exchange term $B_{\v p}$ by a phase factor
which depends on the spins adjacent to the plaquette $\v p$ \cite{LWtoap}. In
the cubic lattice model (\ref{strnetH}), this factor is of the form 
$(-1)^{S^{z}_{5} + S^{z}_{6}}$, where $S_{5}, S_{6}$ are two of the 16 spins 
adjacent to $\v p$. The modified Hamiltonian is thus 
\begin{align}
\label{twistedH}
H &= V\sum_{\v I}  Q^2_{\v I} + g \sum_{\v p} (\tilde{B}_{\v p}+h.c.)
+ \del J \sum_{\v i} (S^z_{\v i})^2 
,
\\
\tilde{B}_{\v p} &= S^+_{1} S^-_{2} S^+_{3} S^-_{4}(-1)^{S^z_5+S^z_6}
,\ \ \ \ \ \
Q_{\v I} = (-1)^{\v I}\sum_{\text{legs of }\v I} S_{\v i}^z
\nonumber 
\end{align}
The two spins $S_5, S_6$ associated with each plaquette $\v p$ are specified as
follows: first, one projects the cubic lattice onto the plane. Then one 
examines the $16$ spins adjacent to each plaquette $\v p$. Two of these spins 
will be located on links that ``cross'' the boundary of the plaquette $\v p$ -- 
in the sense that they cross a closed curve drawn just inside but along the 
boundary of $\v p$. The spins $S_5, S_6$ are precisely the two spins located on
these ``crossed'' links (see Fig. \ref{cubferm}).

The ground state of (\ref{twistedH}) exhibits a different type of string-net
condensation from the previous two models. While string fluctuations still 
correspond to $U(1)$ gauge bosons (e.g. artificial photons), the ends of 
strings now correspond to charged fermions (e.g. artificial electrons).

\section{Are we living in a noodle soup?}

We have seen that string-net condensed states naturally give rise to
gauge bosons (such as photons) and fermions (such as electrons). Thus, the 
existence of photons and electrons is no longer mysterious if we assume that 
our vacuum is a string-net condensate. Photons are vibrations of condensed 
strings, while electrons are the ends of the strings.

But is our vacuum really a string-net condensed state? Photons and electrons
are just two of the elementary particles in nature. So the real question is -- 
can string-net theory explain the other elementary particles? The answer is yes 
and no. String-net condensation naturally explains three of the mysteries of 
nature discussed in the introduction -- identical particles,  gauge interactions 
and Fermi statistics. But so far we do not know how to explain the fourth and the
fifth mysteries -- chiral fermions and gravity. In terms of elementary 
particles, we can construct a string-condensed local bosonic model that 
produces $U(1)$ gauge bosons (photons), $SU(3)$ gauge bosons (gluons), 
leptons (which includes electrons), and quarks \cite{Wqoem}, but we do not 
know how to produce the neutrinos, $SU(2)$ gauge bosons, or gravitons. 

The problem with the neutrinos and the $SU(2)$ gauge bosons is the famous
chiral-fermion problem \cite{L0128}. Neutrinos are chiral fermions and the 
$SU(2)$ gauge bosons couple chirally to other fermions. At the moment, we do 
not know how to obtain chiral fermions and chiral gauge theories from 
\emph{any} local lattice model, much less a local bosonic model. 

Gravity is also a formidable problem. To obtain general relativity from a local
bosonic model, one must develop a quantum theory of gravity, a notoriously
difficult task. However, there is one possible approach: loop quantum gravity
\cite{Smo02}. Remarkably, it appears that the theory of loop quantum gravity
can be reformulated in terms of a particular kind of string-net, where the
strings are labeled by positive integers. \footnote{String-nets with positive
integer labeling were first introduced by Penrose \cite{P71}, and are known
as ``spin networks'' in the loop quantum gravity community. More recently, 
researchers in this field considered the generalization to arbitrary labelings 
\cite{KauL94,Tur94}. These generalized spin networks have the same mathematical 
structure as string-nets. However, we would like to point out that the physical 
meaning of spin networks is fundamentally different from that of string-nets. 
Spin networks are the basic building blocks of loop quantum gravity models. In 
contrast, string-nets describe the pattern of quantum entanglement in the ground 
states of certain spin models. In short, spin networks 
are components of a model while string-nets describe a type of order. The main issue 
in this paper is to find a kind of ordering in spin models that leads to emergent 
photons and electrons. We find that ``particle'' condensation does not work but 
``string'' condensation does work. This is why we introduce the term ``string-net'': 
to stress the stringy character of the ordering.} This means that, in addition to 
gauge interactions and Fermi statistics, string-net condensation in a spin model may 
also give rise to gravity!

\section{Emergence \emph{vs.} reductionism}

In this paper, we propose local bosonic models as a possible origin of elementary
particles. But local bosonic models are far from unique. Should we be 
worried about being overwhelmed with possibilities? If one takes a reductionist 
point of view, this is indeed a serious concern. The string-net picture does not 
tell us how to derive a unique local bosonic model that describes our universe, and 
the models it does suggest are neither simple nor beautiful.

However, according to the point of view of emergence \cite{A7293} -- the 
point of view we take in this paper -- this is not an issue. We are not 
interested in the details of the particular model that produces the observed 
elementary particles. We expect that these details are both complicated and
irrelevant to the low energy emergent physics we observe around us. Instead, we are 
interested in the general mechanism that leads to these particles. In this paper we 
have shown that string-net condensation may be one such mechanism. Photons and 
electrons will emerge if the local bosonic models are in a particular string-net 
condensed phase, irrespective of microscopic details.

On theoretical grounds, string-net condensation appears to be a promising
approach to understanding our universe. Ultimately, however, the validity of
the string-net picture, or more generally the condensed matter picture of the
universe, will be decided by experiment. As we probe nature at shorter and
shorter distance scales, we will either find increasing simplicity, as
predicted by the reductionist particle physics paradigm, or increasing
complexity, as suggested by the condensed matter point of view. We will either
establish that photons and electrons are elementary particles, or we will
discover that they are emergent phenomena -- collective excitations of some
deeper structure that we mistake for empty space.
  
This research is supported by NSF Grant No. DMR--01--23156, NSF-MRSEC Grant No.
DMR--02--13282, and NFSC no. 10228408.


\end{document}